\documentclass[%
 reprint,
 amsmath,amssymb,
 aps,pra,
]{revtex4-2}

\usepackage{graphicx}
\usepackage{graphics}
\usepackage{dcolumn}
\usepackage{bm}

\begin{document}

\title{Thick fluid disks around binary black holes}

\author{S.~V. Chernov}
\affiliation{Astro Space Center, Lebedev Physical Institute, 84/32 Profsoyuznaya street,\\ 117997, Moscow, Russia}
\email{chernov@lpi.ru}

\begin{abstract}
A model of a thick fluid disk around a binary black hole is considered. A binary black hole is described by the Majumdar-Papapetrou  solution. The hydrodynamic equations in this metric are written out. Exact analytical solutions are presented. Generalization to the case of a toroidal magnetic field is carried out.
\end{abstract}

\maketitle

\section{Introduction}

Accretion disks around single black holes have important astrophysical significance. On the other hand, On September 14, 2015, the Laser Interferometer Gravitational-Wave Observatory detected a gravitational-wave signal from the source GW150914 - thereby confirming the existence of binary black holes \cite{abbott}. The masses of such black holes are small, only a few tens of solar masses. In addition, binary supermassive black holes can exist in the center of active galactic nuclei \cite{sudou}. One of the main examples is the OJ 287 system. It is believed that this system is a binary black hole with masses of 18 billion and 125 million solar masses \cite{titarchuk}. The orbital period of such a system is 12 years and it is assumed that an accretion disk rotates around a more massive black hole, which is penetrated by a less massive black hole. Such systems can be formed due to the merger of two or more galaxies with massive black holes \cite{sudou}. Thus, accretion disks may also exist around binary black holes, the study of which is of astrophysical interest.

In this paper, thick accretion disks around binary black holes are investigated. For simplicity, it is assumed that double black holes are described by the Majumdar-Papapetrou metric. Chapter 2 describes the metric and its main properties, which are used in this work. In Chapter 3, the hydrodynamic equation in this metric is written out and exact analytical solutions are constructed. In chapter 4, a generalization is constructed for the case of a toroidal magnetic field. Chapter 5 provides a conclusion.

In the paper, we use the geometrical units, G=c=1.

\section{Basic equations}

In this paper we will consider the simplified model of a binary black hole, which is described by the Majumdar-Papapetrou solution \cite{majumdar,papapetrou}.
The solution has the form \cite{majumdar,papapetrou}
\begin{eqnarray}
 ds^2=-\Omega^{-2}dt^2+\Omega^2(dx^2+dy^2+dz^2),
 \label{metricDekart}
\end{eqnarray}
where $\Omega=1+\sum\limits_{i}\frac{m_i}{r_i}$, "i" is a number of black holes,
$$r_i=\sqrt{(x-x_i)^2+(y-y_i)^2+(z-z_i)^2},$$ $x_i,y_i,z_i$ - are the coordinates location and $m_i$ - is the mass of black holes. For the case when one black hole is considered, this metric passes into the metric of the extreme Reissner-Nordstrom black hole. Here we consider the case of two black hole, when $i=2$ and without loss of generality, we will assume that black holes are located on the z axis with coordinates (0,0,1) and (0,0,-1) respectively. 

Let's rewrite the metric (\ref{metricDekart}) in a cylindrical coordinate system. To do this, we introduce cylindrical coordinates
\begin{eqnarray}
 x&=&r\cos\phi,\nonumber\\
 y&=&r\sin\phi,\nonumber\\
 z&=&z,
\end{eqnarray}
in which the metric (\ref{metricDekart}) will be written as
\begin{eqnarray}
 ds^2=-\Omega^{-2}dt^2+\Omega^2(dr^2+r^2d\phi^2+dz^2),
 \label{metriccylindr}
\end{eqnarray}
where $\Omega=1+\sum\limits_{i=1}^2\frac{m_i}{r_i}$, $r_i=\sqrt{r^2+(z-z_i)^2}$. This metric has two Killing vectors, $\xi_{(t)}=\partial/\partial t$ and $\xi_{(\phi)}=\partial/\partial\phi$ which generate time shifts and rotations around the axis of symmetry $z$. 

We will also need a metric determinant that is equal to
$\sqrt{-g}=r\Omega^2$.

\section{solutions of hydrodynamic equations}

In this section we consider a perfect fluid in a stationary, axisymmetric binary black holes where the self-gravity of the fluid is ignored. This means that the following conditions are satisfied:
$\partial/\partial t=0$ and $\partial/\partial\phi=0$ and all values depend on two variables, r and z. 
It is also assumed that the fluid rotates only around the axis of symmetry z. This means that the radial and z components of the four-velocities are equal to zero, $u^z=u^r=0$ and only the temporal and azimuthal components are different from zero, $u^t\neq0$, $u^\phi\neq0$. It is not difficult to see that under such conditions the continuity equation is performed identically. The four-velocity of the fluid satisfies the condition, $g_{\alpha\beta}u^\alpha u^\beta=-1$, which in our case will be rewritten as
\begin{eqnarray}
 u^{t2}-r^2\Omega^4u^{\phi2}=\Omega^2.
 \label{utuphi}
\end{eqnarray}
From the conservation laws $T^{\alpha\beta}_{\,\,\,\,\,;\beta}=0$ for the energy-momentum tensor of an ideal fluid
\begin{eqnarray}
 T^{\alpha\beta}=(P+\rho)u^\alpha u^\beta+Pg^{\alpha\beta}
\end{eqnarray}
only two components (r,z) will remain nonzero, which can be written as
\begin{eqnarray}
 \frac{1}{p+\rho}\frac{\partial P}{\partial a}=\frac{1}{\Omega}\frac{\partial\Omega}{\partial a}+\frac{u^{\phi2}}{2\Omega^2}\frac{\partial}{\partial a}(r^2\Omega^4),
 \label{ldpda}
\end{eqnarray}
where notation "a" takes two values $a=r$ or $z$ and $P$ - is the pressure, $\rho$ - is the energy density. 

To integrate equation (\ref{ldpda}), we need to express the pressure and energy density in terms of enthalpy, $h$. Assuming that the entropy, $s$, is constant along the fluid flow line, then it is easy to obtain that
\begin{eqnarray}
 \frac{dP}{P+\rho}=\frac{dh}{h}.
 \label{dpdh}
\end{eqnarray}
Thus, we have a set of equations (\ref{utuphi}) and (\ref{ldpda}), for the solution of which we need to impose additional constrain (see, \cite{fishbone,kozlowski}).

\subsection{Fishbone-Moncrief solution}
\label{ch-fish}

\begin{figure}
\includegraphics[scale=0.5]{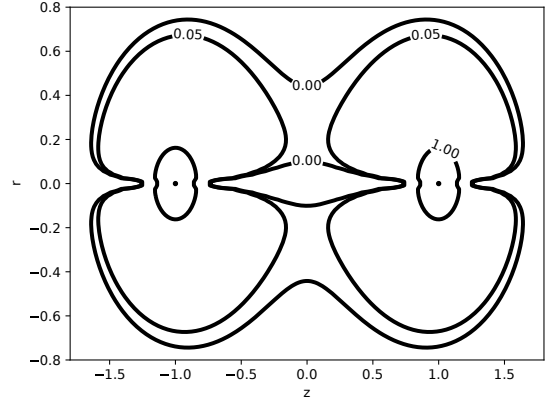}
\caption{Contours $\ln(h)=0;0.05;1$ are shown by solid curves. Black holes are located at (0,-1) and (0,1). The masses of black holes are equal, $m_1=1$ and $m_2=1$.}
\label{fig1}
\end{figure}

\begin{figure}
\includegraphics[scale=0.5]{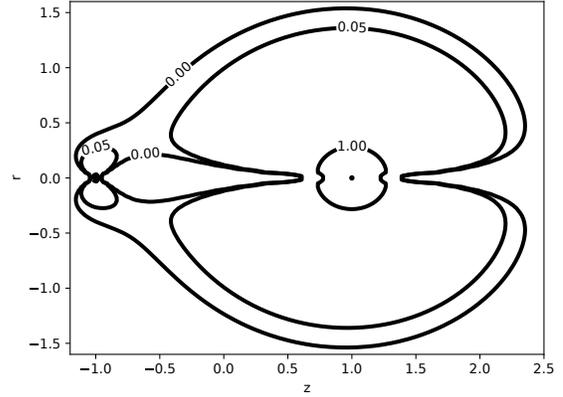}
\caption{Contours $\ln(h)=0;0.05;1$ are shown by solid curves. Black holes are located at (0,-1) and (0,1). The masses of black holes are equal, $m_1=1$ and $m_2=0.1$, respectively.}
\label{fig2}
\end{figure}

One of such constrain was proposed in the work \cite{fishbone}. They assumed that the value $l=u_\phi u^t$ remains constant. Then using (\ref{utuphi}) you can get
\begin{eqnarray}
 u^{\phi2}=\frac{-1+\sqrt{1+\frac{4l^2}{r^2\Omega^4}}}{2r^2\Omega^2}
 \label{uphi2fishbone}
\end{eqnarray}
and rewrite the equation (\ref{ldpda}) in the form
\begin{eqnarray}
 \frac{1}{h}\frac{\partial h}{\partial a}=\frac{1}{\Omega}\frac{\partial\Omega}{\partial a}+\frac{-1+\sqrt{1+\frac{4l^2}{r^2\Omega^4}}}{4r^2\Omega^4}\frac{\partial}{\partial a}(r^2\Omega^4).
 \label{dhda}
 \label{h}
\end{eqnarray}
It is easy to integrate the above equation (\ref{dhda}). As a result, we get
\begin{eqnarray}
 \ln(h)=\frac{1}{4}\ln\left(1+\frac{2l^2}{r^2\Omega^4}+\sqrt{1+\frac{4l^2}{r^2\Omega^4}}\right)+\nonumber\\
 +\ln\Omega-\frac{1}{2}\sqrt{1+\frac{4l^2}{r^2\Omega^4}}-\ln(h_{c}),
 \label{fish-fluid-sol}
\end{eqnarray}
where it is convenient to determine the integration constant in the plane (z=0) perpendicular to the axis connecting the black holes in the point ($z=0$, $r=r_b$)
\begin{eqnarray}
 \ln(h_{c})=\ln\Omega_{z=0,r=r_b}-\frac{1}{2}\sqrt{1+\frac{4l^2}{r_b^2\Omega^4_{z=0,r=r_b}}}+\nonumber\\
 +\frac{1}{4}\ln\left(1+\frac{2l^2}{r_b^2\Omega^4_{z=0,r=r_b}}+\sqrt{1+\frac{4l^2}{r_b^2\Omega^4_{z=0,r=r_b}}}\right).
\end{eqnarray}
This solution (\ref{fish-fluid-sol}) is a general solution of the equations (\ref{utuphi}) and (\ref{ldpda}). The figures (\ref{fig1}) and (\ref{fig2}) show examples of the contours of the logarithm of enthalpy $\ln(h)$ for a disk around two black holes. In these figures, negative radii correspond to azimuthal angles $\phi+\pi$. The boundary conditions were determined at the point ($z=0$, $r_b=0.1$). The constant value $l$ was chosen equal to $l^2=0.1$.

\subsection{Kozlowski et.al. solution}

Another constrain was suggested in the paper \cite{kozlowski}. They assumed that the value $l=-u_\phi/u_t$ remains constant. Then it follows from the equation (\ref{utuphi}) that $\phi$-component of the four-velocity is equal to
\begin{eqnarray}
 u^{\phi2}=\frac{l^2}{r^2\Omega^2(r^2\Omega^4-l^2)}.
 \label{uphi2kozlo}
\end{eqnarray}
It follows from this expression (\ref{uphi2kozlo}) that the condition $r^2\Omega^4>l^2$ must be fulfilled, which cannot be fulfilled near the points $r\approx0$, $z\neq\pm1$. Therefore, in this case, the solution will not completely cover the space-time around the binary black hole. Formally, in this case, we can also fully integrate these equations, (\ref{ldpda}). Using the equations (\ref{dpdh}) and (\ref{uphi2kozlo}), we can rewrite the equation (\ref{ldpda}) as
\begin{eqnarray}
 \frac{1}{h}\frac{\partial h}{\partial a}=\frac{1}{\Omega}\frac{\partial\Omega}{\partial a}+\frac{l^2}{2r^2\Omega^4(r^2\Omega^4-l^2)}\frac{\partial}{\partial a}(r^2\Omega^4).
\end{eqnarray}
After integration we get
\begin{eqnarray}
 \ln(h)=\ln\Omega+\frac{1}{2}\ln\left(1-\frac{l^2}{r^2\Omega^4}\right)-\ln(h_{c}).
\end{eqnarray}
where the integration constant is defined in the same way as in the previous case
\begin{eqnarray}
 \ln(h_{c})=\ln\Omega_{z=0,r=r_b}+\frac{1}{2}\ln\left(1-\frac{l^2}{r^2_b\Omega^4_{z=0,r=r_b}}\right).
\end{eqnarray}

\subsection{Another exact solution}
\label{anothsolfluid}

\begin{figure}
\includegraphics[scale=0.5]{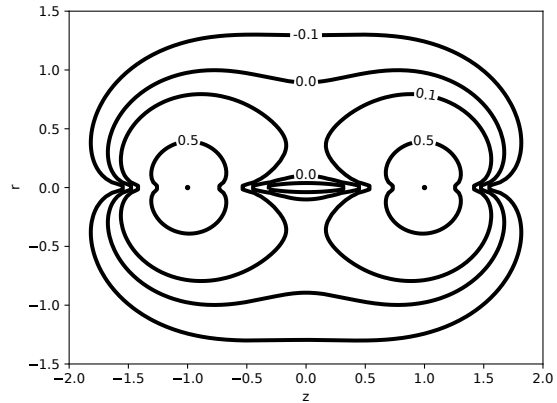}
\caption{Contours of the logarithm of the constant enthalpy $\ln(h)=-0.1;0;0.1;0.5$ are shown by solid curves. Black holes are located at (0,-1) and (0,1). The mass of black holes are equal, $m_1=m_2=1$.}
\label{fig3}
\end{figure}

\begin{figure}
\includegraphics[scale=0.5]{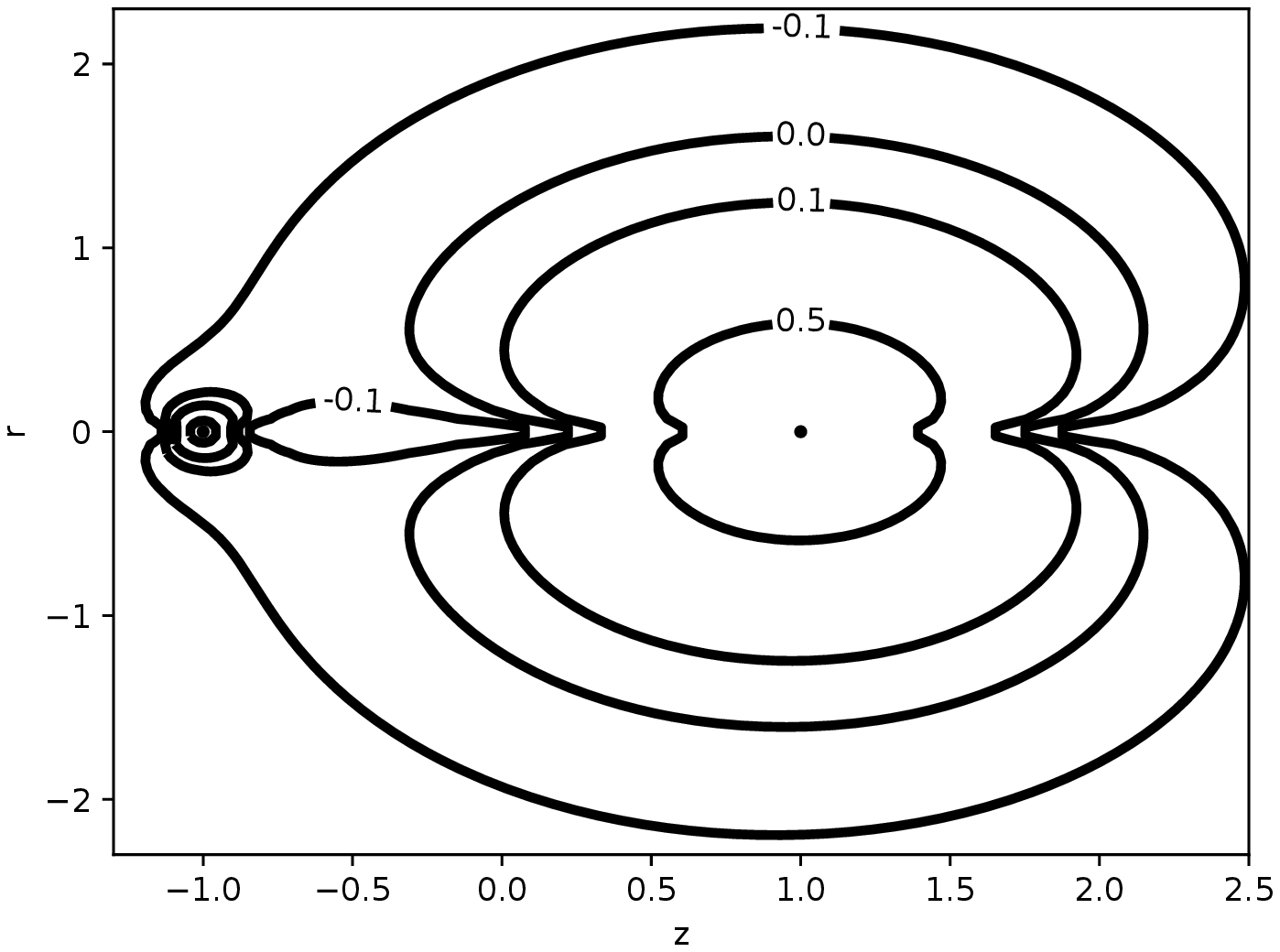}
\caption{Contours of the logarithm of the constant enthalpy $\ln(h)=-0.1;0;0.1;0.5$ are shown by solid curves. Black holes are located at (0,-1) and (0,1). The mass of black holes are equal, $m_1=1$ and $m_2=0.1$.}
\label{fig4}
\end{figure}

By making various assumptions, other exact analytical solutions can be obtained. For example, assuming that the value $l=u_\phi u^\phi$ remains constant, using expressions (\ref{ldpda}) and (\ref{dpdh}) we obtain the desired equation
\begin{eqnarray}
 \frac{1}{h}\frac{\partial h}{\partial a}=\frac{1}{\Omega}\frac{\partial\Omega}{\partial a}+\frac{l}{2r^2\Omega^4}\frac{\partial}{\partial a}(r^2\Omega^4).
\end{eqnarray}
After integration, we get an exact analytical solution
\begin{eqnarray}
 \ln(h)=\ln\Omega+\frac{l}{2}\ln(r^2\Omega^4)-\ln(h_c)
\end{eqnarray}
with the integration constant equal to
\begin{eqnarray}
 \ln(h_c)=\ln\Omega_{z=0,r=r_b}+\frac{l}{2}\ln(r_b^2\Omega^4_{z=0,r=r_b}).
\end{eqnarray}
The figures (\ref{fig3}) and (\ref{fig4}) show examples of the distribution of the contours of the constant enthalpy $\ln(h)=-0.1;0;0.1;0.5$ for various parameters of the problem. In the figure (\ref{fig3}), the mass of black holes is equal to, $m_1=m_2=1$, and in the figure (\ref{fig4}) the mass of black holes is equal to, $m_1=1$, $m_2=0.1$. The value $l$ was chosen equal to, $l=0.1$.

\section{solutions with a toroidal magnetic field}

Following the work of \cite{komissarov}, we can generalize the above solutions to the case of the presence of a toroidal magnetic field. A set of ideal relativistic MHD equations are
\begin{eqnarray}
 g_{\alpha\beta}u^\alpha u^\beta=-1,\nonumber\\
 T^{\alpha\beta}_{\,\,\,\,\,;\beta}=0,\nonumber\\
 (nu^\alpha)_{;\alpha}=0,\nonumber\\
 F_{\mu\nu,\lambda}+F_{\nu\lambda,\mu}+F_{\lambda\mu,\nu}=0,
 \label{setmhd}
\end{eqnarray}
where $n$ is the baryon number density, $F_{\mu\nu}$ is the electromagnetic field tensor and
\begin{eqnarray}
 T^{\alpha\beta}=(P+\rho+b^2)u^\alpha u^\beta+(P+\frac{b^2}{2})g^{\alpha\beta}-b^\alpha b^\beta,
 \label{mhdtenzor}
\end{eqnarray}
where $b^\alpha$ is the four-vector of magnetic field \cite{komissarov}. Also it's suppose that the flow is stationary, $\partial/\partial t$,  and axisymmetric, $\partial/\partial\phi$, the velocity of fluid and magnetic field have only toroidal component, $u^r=u^\theta=0$, $b^r=b^\theta=0$. Then the third and forth equations (\ref{setmhd}) are satisfied indentically. The r and z component of second eqution (\ref{setmhd}) are rewritting in the form
\begin{eqnarray}
 \frac{1}{\sqrt{-g}}\frac{\partial}{\partial a}\left(\sqrt{-g}\left(p+\frac{1}{2}b^2\right)\right)=\frac{1}{2}g_{\gamma\delta,a}T^{\gamma\delta}.
 \label{qwe1}
\end{eqnarray}
If we substitute metric coefficients (\ref{metriccylindr}) and the energy-momentum tensor (\ref{mhdtenzor}) into the equation (\ref{qwe1}), then after tedious transformations we can get an expression of the form
\begin{eqnarray}
 \frac{\partial}{\partial a}(p+\frac{b^2}{2})=
 \frac{(p+\rho+b^2)u^{t2}-b^{t2}}{\Omega^3}\frac{\partial\Omega}{\partial a}+\nonumber\\
 +\frac{(p+\rho+b^2)u^{\phi2}-b^{\phi2}}{2}\frac{\partial}{\partial a}(r^2\Omega^2).
 \label{dpdamhd}
\end{eqnarray}

In order to integrate this equation, it is necessary to have an equation of state $p=k\rho^\gamma$ and to associate enthalpy $P+\rho$ with magnetic pressure $b^2$. Following the work \cite{komissarov}, we assume that these quantities are connected in a linear way of the form
\begin{eqnarray}
 \beta=\frac{P+\rho}{b^2}.
\end{eqnarray}
For simplicity, let's assume that the parameter, $\beta$, is a constant. 
Using the parameters $\beta$, we obtain a final equation that describes the distribution of a fluid with a toroidal magnetic field around binary black holes.
\begin{eqnarray}
\frac{\beta}{1+\beta}\frac{1}{p+\rho}\frac{\partial p}{\partial a}+\frac{1}{2(1+\beta)b^2}\frac{\partial b^2}{\partial a}=\frac{1}{\Omega}\frac{\partial\Omega}{\partial a}+\nonumber\\
+\frac{u^{\phi2}}{2\Omega^2}\frac{\beta}{1+\beta}\frac{\partial}{\partial a}(r^2\Omega^4)
 -\frac{1}{2}\frac{1}{1+\beta}\frac{1}{r^2\Omega^2}\frac{\partial}{\partial a}(r^2\Omega^2).
 \label{finalmhdeq}
\end{eqnarray}
We explicitly integrate this equation below by applying various constrains to the four-velocity of the fluid.

\subsection{Fishbone-Moncrief solution}

\begin{figure}
\includegraphics[scale=0.5]{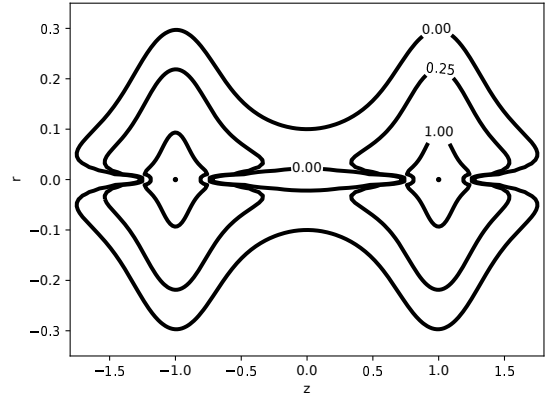}
\caption{Contours of the left side of the equation (\ref{fish-mhd-sol}) for values $0;0.25;1.0$ are shown by solid curves. Black holes are located at (0,-1) and (0,1). The mass of black holes are equal, $m_1=m_2=1.0$.}
\label{fig5}
\end{figure}

Assuming that the constraints of the form, $l=u_\phi u^t$, as in the work \cite{fishbone} are fulfilled and using the equation (\ref{uphi2fishbone}) from equation (\ref{finalmhdeq}), we obtain an exact analytical solution for the distribution of fluid in a toroidal magnetic field around a double black hole.
\begin{eqnarray}
 \frac{\beta}{1+\beta}\frac{\gamma}{\gamma-1}\ln(1+k\rho^{\gamma-1})+\frac{1}{2(1+\beta)}\ln(b^2)=\nonumber\\
 =\ln(\Omega)+\frac{\beta}{4(1+\beta)}\ln(1+\frac{2l^2}{r^2\Omega^4}+\sqrt{1+\frac{4l^2}{r^2\Omega^4}})-\nonumber\\
 -\frac{\beta}{2(1+\beta)}\sqrt{1+\frac{4l^2}{r^2\Omega^4}}-\frac{\ln(r^2\Omega^2)}{2(1+\beta)}-\ln(h_{c}).
 \label{fish-mhd-sol}
\end{eqnarray}
The integration constant is defined at a point ($z=0$, $r=r_{b}$) in the same way as in the chapter (\ref{ch-fish}). In the limiting case, when the parameter $\beta\rightarrow\infty$ tends to infinity, we get the solution (\ref{fish-fluid-sol}). The figure (\ref{fig5}) show examples of contours of the left side of the equation (\ref{fish-mhd-sol}) for values $-0.2;-0.1;0.25$ for the following parameters: $\beta=2$, $m_1=m_2=1$, $l^2=0.1$, $r_b=0.1$.

\begin{figure}
\includegraphics[scale=0.5]{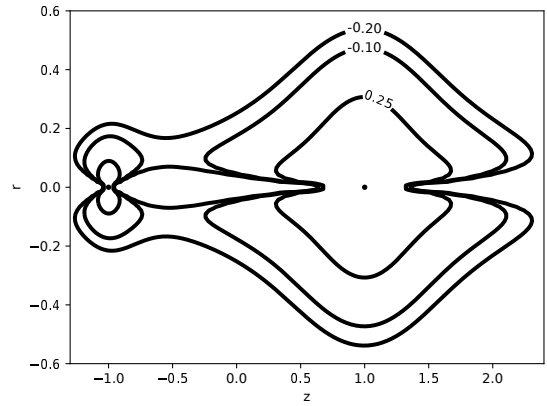}
\caption{Contours of the left side of the equation (\ref{fish-mhd-sol}) for values $-0.2;-0.1;0.25$ are shown by solid curves. Black holes are located at (0,-1) and (0,1). The mass of black holes are equal, $m_1=1.0$ and $m_2=0.1$.}
\label{fig6}
\end{figure}

If we take the mass of the second black hole equal $m_2=0.1$, then the contours of the left side of the equation for the same parameters of the problem, $m_1=1$, $l^2=0.1$, $r_b=0.1$ will change and take the form as in the figure (\ref{fig6}).

Comparing Figures (\ref{fig1}) and (\ref{fig5}) or (\ref{fig2}) and (\ref{fig6}), it can be seen that the toroidal magnetic field can greatly change the behavior of the contours and the qualitative picture of the distribution of fluid near binary black holes.

\subsection{Another exact solution}

Finally, consider the case when the constrain is described by the relation $l=u_\phi u^\phi$, see (\ref{anothsolfluid}). Then the solution of the equation (\ref{finalmhdeq}) is easy to obtain. As a result, we get
\begin{eqnarray}
 \frac{\beta}{1+\beta}\frac{\gamma}{\gamma-1}\ln(1+k\rho^{\gamma-1})+\frac{1}{2(1+\beta)}\ln(b^2)=\nonumber\\
 =\ln(\Omega)+\frac{\beta l}{1+\beta}\ln(r\Omega^2)-\frac{\ln(r\Omega)}{1+\beta}-\ln(h_c),
 \label{anoth-mhd-sol}
\end{eqnarray}
where the integration constant, $\ln(h_c)$, is defined similarly to the previous cases.

\begin{figure}
\includegraphics[scale=0.5]{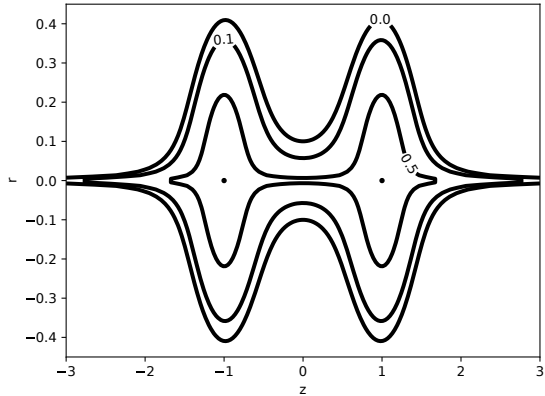}
\caption{Contours of the left side of the equation (\ref{anoth-mhd-sol}) for values $0;0.1;0.5$ are shown by solid curves. Black holes are located at (-1,0) and (1,0). The mass of black holes are equal, $m_1=m_2=1.0$. $\beta=3$}
\label{fig7}
\end{figure}

\begin{figure}
\includegraphics[scale=0.5]{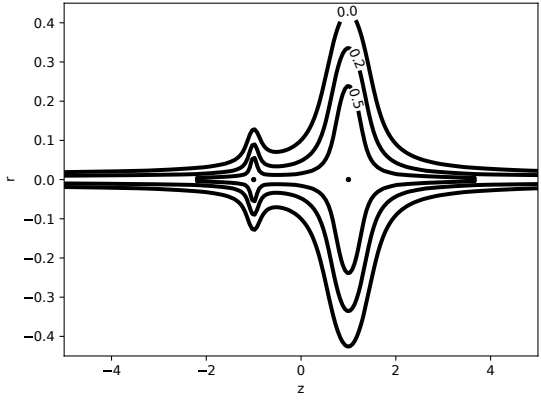}
\caption{Contours of the left side of the equation (\ref{anoth-mhd-sol}) for values $0;0.2;0.5$ are shown by solid curves. Black holes are located at (-1,0) and (1,0). The mass of black holes are equal, $m_1=1.0$ and $m_2=0.1$. $\beta=2$}
\label{fig8}
\end{figure}

In the figures (\ref{fig7}) and (\ref{fig8}) you can see the contours of the left side of the equation (\ref{anoth-mhd-sol}) for the following parameters: $m_1=m_2=1$, $l=0.1$ $\beta=3$ - for fig. (\ref{fig7}) and $m_1=1$, $m_2=0.1$, $l=0.1$ $\beta=2$ for fig. (\ref{fig8}). Comparing Figures (\ref{fig3}) (\ref{fig4}) with Figures (\ref{fig7}) (\ref{fig8}), one can see a strong difference associated with the absence of a thick disk between black holes in the presence of a strong toroidal magnetic field.

\section{Conclusions}

In this paper, we generalized the relativistic theory of thick accretion disks to the case of binary black holes. Black holes were described by the Majumdar-Papapetrou solution, in which the mass of a black hole is compensated by an electric charge. A method for constructing a thick accretion disk in the hydrodynamic case and with the presence of a toroidal magnetic field was described. Exact analytical solutions are written out. These solutions can be used in numerical MHD calculations of the evolution of thick disks around binary black holes as the initial conditions of the problem.


\end{document}